\documentstyle[aps,prc,twocolumn,epsf,floats]{revtex}

\newcommand{\Tr}{{\mbox{Tr}}}
\newcommand{\bzero}{{\bf 0}}
\newcommand{\bx}{{\bf x}}
\newcommand{\by}{{\bf y}}
\newcommand{\br}{{\bf r}}
\newcommand{\bR}{{\bf R}}
\newcommand{\bQ}{{\bf Q}}
\newcommand{\Real}{{\rm Re}}

\begin{document}
\draft

\twocolumn[\hsize\textwidth\columnwidth\hsize\csname @twocolumnfalse\endcsname
\title{
Penetration of a high energy $Q \bar Q$ bound state through SU($N$) color background}
\author{H.\ Fujii}
\address{Institute of Physics, University of Tokyo
3-8-1 Komaba, Meguro, Tokyo 153-8902 }
\date{May 24, 2002}
\maketitle
\begin{abstract}
Attenuation of the singlet quark--anti-quark bound state
passing through a random external SU($N$) color field at high energy
is studied within the eikonal framework as a model for the absorption 
of a quarkonium by nuclear target.
For a thick target of length $L$ the 
penetration probability is suppressed non-exponentially and
behaves asymptotically $\propto 1/LN^2$ with $1/N^2$ indicating the 
equi-partitioning in the color space.
An approximate rational expression for the penetration of the small
Gaussian bound state is found which allows the explicit $L$ and $N$
dependences, and shows the attenuation is in the powers of $L$
although the approach to the asymptote is rather slow.
\end{abstract}
]

\section{Introduction}



Color dipole is a simplest physical (color singlet) state of QCD,
and hence may be used as a tool
to study more complex objects in QCD. In Nature
it is realized as heavy quarkonia ($c\bar c$ or $b \bar b$),
a virtual $q \bar q$ pair in deeply inelastic scattering seen
in the target rest frame, and so on.
In the context of the nucleus-nucleus ($AB$) collision physics,
the modification of the quarkonium productions in the QCD media has
been extensively studied 
both experimentally and theoretically\cite{WJ98}
since the proposal as a signal of the QCD plasma formation\cite{MS86}.

In the previous paper\cite{FM02}
 we studied the penetration of a high energy
$Q\bar Q$ bound state traversing a
nuclear target, which is modeled as soft random classical fluctuations of
SU(2) gauge fields. 
Working in the eikonal framework we showed that the
penetration probability of the color singlet bound state
attenuates asymptotically in $1/L$ with $L$ the target thickness
\cite{ZKL81,Ba96},
not in the exponential form. This is essentially the same effect 
as the so-called {\it superpenetration}, conjectured theoretically for the
ultra-relativistic positronium passing through a metallic
foil\cite{LP81,N81,Ba96}.

The aim of this paper is to extend our formulation to general SU($N$) case.
We explicitly show the $1/N^2 L$ suppression 
of the penetration probability of the singlet bound state
in the asymptotic region. An analytic, approximate formula which
well reproduces the $N$- and $L$-dependences is given for 
the Gaussian bound state and shows that the penetration
is power-suppressed by $1/L$ although the approach to 
the asymptotic form is rather slow.

Within the eikonal approximation we deal with the multiple scatterings of
the quark and anti-quark with
the phase factors accumulated along the classical straight paths
with fixed transverse positions.
The bound state wavefunction described in terms of these phases should
generally contain various excited states of the pair
from the initial state. In this sense our treatment
of the pair propagation on the background field
includes the quantum-mechanical coherence. 
The random phase disarranges the wavefunction to reduce
the overlap with the initial state on average.

In Sec.~\ref{sec:formula} the basic eikonal formulas for SU($N$) theories
are given, 
and Sec.~\ref{sec:target} defines the target average following
Ref.\cite{KW01}. Our main result on the survival probability
of the singlet bound state is presented in
Sec.~\ref{sec:penetration}. Secs.~\ref{sec:discussion} and \ref{sec:summary}
are devoted to discussions and summary. In Appendix \ref{app1} the
target averaging of the pair eikonal factors is  briefly explained.
Our framework allows an analytic solution to the SU(2) model 
with the finite gauge fluctuation amplitude of Ref.\cite{HLN90},
which is given in Appendix~\ref{app2}.

\section{Eikonal approximation for pair propagation}
\label{sec:formula}

Let us consider the quark--anti-quark bound state projected onto a
large target nucleus of thickness $L$ at very high energies. 
We choose the target rest frame in which the projectile bound state
travels with ultra-relativistic momentum $P$.
In such a high energy process the deflections of the constituents of
the bound state can be ignored and their transverse positions are frozen
during the interaction with the target.
The eikonal approximation for the propagation of these constituents 
is our basic framework to describe the interactions of the bound state
with the soft classical gluon fields created by fluctuations in the
nucleons in the target nucleus\cite{N91,B96,BH96,H00}. 

The bound state is described by a wavefunction
$\varphi(\br,z)\xi$ where $\br=\bx-\by$ is the relative transverse
position of the quark at $\bx$ and anti-quark $\by$, 
$z$ is the fraction of the longitudinal momentum carried by the quark, and
$\xi$ is the color wavefunction of the pair.
Except for the vicinity of the endpoints, $z=0$ and 1,
the longitudinal momenta of the constituents are large enough
that the quark and anti-quark are regarded as moving
along the classical, straight-line trajectories.
The interaction with the external gauge fields is described
by the usual eikonal quark-gluon vertex $igv_\mu A^\mu(x)$ with 
$v_\mu=p_\mu/E_p$ and $A_\mu=A_\mu^a t^a$ ($t^a$ are the SU($N$)
generators normalized as $\Tr t^a t^b=\frac{1}{2}\delta^{ab}$).
Traveling through the target nucleus, the quark acquires the
eikonal phase, or Wilson line, $W(\bx)$ accumulated along the path as
\begin{equation}
W(\bx;A)=
{\cal P}e^{ig\int_{-\infty}^{\infty} d x^+ A^{a-}(\bx,x^+)t^a},
\end{equation}
where we have taken the limit $v^\mu=(1,0,0,1)$ and 
introduced the notations 
$x^\pm =\frac{1}{\sqrt{2}}(t\pm z)$ and
$A^\pm =\frac{1}{\sqrt{2}}(A^0\pm A^3)$. This is a matrix in the color space
and ${\cal P}$ indicates the path ordering
of the product. 
The corresponding
eikonal factor for the anti-quark is simply
$W^\dagger(\by;A)=
{\cal P}e^{ig\int_{+\infty}^{-\infty} dy^+ A^{a-}(\by,y^+)t^a}$
since it belongs to the color representation conjugate to the quarks.
After the interaction with the target nucleus, hence,
the quark--anti-quark pair gets the total eikonal phase
\begin{eqnarray}
&&U(\bx,\by;A)
=W(\bx;A)W^\dagger(\by;A)  .
\label{eq:eik}
\end{eqnarray}
The scattering amplitude from the initial state 
$\phi_{\rm i}(\bx,\by,z)=\varphi_{\rm i}(\br,z)\xi_{\rm i}$
to the final state 
$\phi_{\rm f}(\bx,\by,z)
=e^{-i\bQ \bR}\varphi_{\rm f}(\br,z)\xi_{\rm f}$ 
with the CM transverse momentum $\bQ$ and 
$\bR=\frac{1}{2}(\bx+\by)$,
is expressed as
\begin{eqnarray}
    \langle \Phi_{\rm f} \phi_{\rm f} |
       U|\phi_{\rm i} \Phi_{\rm i}\rangle
&=&   \int DAd \bx d \by dz 
      \Phi_{\rm f}^*[A]
      e^{i\bQ \bR} \varphi_{\rm f}^*(\br,z)\xi_{\rm f}^\dagger
\nonumber \\
&&
\times
     U(\bx,\by;A)\Phi_{\rm i}[A]
     \varphi_{\rm i}(\br,z)\xi_{\rm i},
\label{eq:amp}
\end{eqnarray}
where $\Phi_{\rm i(f)}$ is the initial (final) state of the target
which  we treat as a functional of soft external fields, $A$.

In studying the penetration probability of the initial bound
state, only the out-going state of the pair is of relevance.
Taking the closure of the final target states, we have
\begin{eqnarray}
&&
\overline{
      \langle \phi_{\rm i} |U^\dagger|\phi_{\rm f} \rangle
      \langle \phi_{\rm f} |U|\phi_{\rm i} \rangle 
}
\equiv
    \langle \Phi_{\rm i} |\ 
      \langle \phi_{\rm i} |U^\dagger|\phi_{\rm f} \rangle
      \langle \phi_{\rm f} |U|\phi_{\rm i} \rangle 
      \ |\Phi_{\rm i}\rangle
\nonumber \\
&=&
      \int DA \
      |\Phi_{\rm i}[A]|^2 \
      \langle \phi_{\rm i} |U^\dagger[A]|\phi_{\rm f} \rangle
      \langle \phi_{\rm f} |U[A]|\phi_{\rm i} \rangle .
\label{eq:average}
\end{eqnarray}
In this form we first calculate the scattering probability of the pair
on a certain background field configuration, 
and then take average over the gauge
configuration distribution $|\Phi_{\rm i}[A]|^2$.
It is more convenient, however, to average over the $A$ configurations
before taking the matrix element with respect to the pair bound state,
because the $A$ fields appear only in the eikonal
factors which are independent of the in- and out-states. 
More explicitly it is written as
\begin{eqnarray}
&&
\overline{
      \langle \phi_{\rm i} |U^\dagger|\phi_{\rm f} \rangle
      \langle \phi_{\rm f} |U|\phi_{\rm i} \rangle 
}
=
      \int d\bx d\by dz d\bar \bx d\bar \by d\bar z
      e^{i\bQ(\bR-\bar \bR)}
\nonumber \\
&&\qquad \times
      \varphi_{\rm f}(\br,z)\varphi_{\rm f}(\bar \br,\bar z)
\overline{P_{\rm fi}(\bx,\by;\bar \bx, \bar \by)}
      \varphi_{\rm i}(\br,z)\varphi_{\rm i}(\bar \br,\bar z)
\ ,
\end{eqnarray}
where $\bar \bx$, $\bar \by$  and $\bar z$ are the coordinates in
the complex conjugate amplitude, and
we introduced a shorthand notation for the product of the eikonal factors,
\begin{equation}
P_{\rm fi}(\bx,\by;\bar \bx, \bar \by)
\equiv
    \xi_{\rm f}^\dagger U        (\bx,\by)\xi_{\rm i}
\   (\xi_{\rm f}^\dagger U(\bar \bx,\bar\by) \xi_{\rm i})^*
\ .
\end{equation}
$P_{\rm fi}$ includes all the dependence of the probability on the
background gauge field, and the probability is obtained as an
integral with the kernel $\overline{P_{\rm fi}}$.

\begin{figure}[tb]
\begin{center}
\epsfxsize=0.3\textwidth
\epsffile{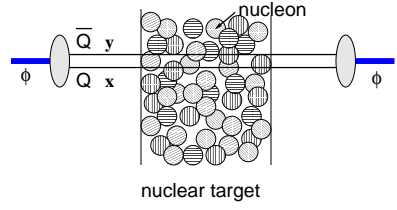}
\caption{Schematic picture of the high energy $Q \bar Q$ pair passing
  through a nuclear target.\label{fig:1}
}
\end{center}
\end{figure}

\subsection{eikonal factors in color irreducible basis }

The expressions for the eikonal factors are much simplified
in the color basis of
the irreducible representations. Since the color singlet and
adjoint states are constructed as
$\xi_{\rm s}=\frac{1}{\sqrt{N}} \sum_{i=1}^N|i \rangle{\mathbf 1} \langle i|
$ and $
\xi_{\rm a}= \sqrt{2} \sum_{i,j=1}^N|i\rangle t^a_{ij}\langle j| ,
$
respectively, out of the quark $|i\rangle$ and the anti-quark $\langle j|$,
the elements of the eikonal factor with the definite color yield 
\begin{eqnarray}
U_{\rm ss}(\bx,\by) &=&
\frac{1}{N}
\Tr(W(\bx)W^\dagger(\by)),
\nonumber \\
U^a_{\rm as}(\bx,\by) &=&
\sqrt{\frac{2}{N}}
\Tr(W(\bx) W^\dagger(\by) t^a),
\nonumber \\
U^a_{\rm sa}(\bx,\by) &=&
\sqrt{\frac{2}{N}}
\Tr(W(\bx) t^a W^\dagger(\by)),
\nonumber \\
U^{ba}_{\rm aa}(\bx,\by) &=&
2\Tr( t^b W(\bx) t^a W^\dagger(\by)).
\label{eq:coloramp}
\end{eqnarray}

These eikonal factors show the crucial features
of the interactions of the pair.
{\it Color Transparency}\cite{BBGG81,BM88}:
An elementary color-singlet dipole
with a vanishing size, $\bx=\by$, does not interact with the gauge
fields at all. 
Actually the quark and the anti-quark acquire the same eikonal phase, and hence
$U_{\rm ss}=1$ and $U_{\rm as}=0$, irrespective of the gauge
fields.
{\it Color Rotation}: An elementary color-adjoint dipole
changes its color orientation as well as the CM motion through 
the interaction with the gauge fields.

This is an $N^2 \times N^2$ matrix in the color space. 
In Ref.~\cite{FM02} we explicitly 
dealt with this full matrix for SU(2) case, and confirmed 
that the color rotation within the adjoint representation is
secondary for the dynamics of the penetration of the color singlet
state. Furthermore the orientation of the color adjoint vector has
almost no meaning in the color gauge theory.
Hence we will concentrate on the transition processes between
the irreducible color representations of the singlet and the adjoint.

Summing up the color orientations, $a=1,\cdots, N^2-1$,
we have the eikonal factors
squared for the singlet-singlet ($P_{\rm ss}$), singlet--adjoint
($P_{\rm as}$ and $P_{\rm sa}$), adjoint--adjoint ($P_{\rm aa}$) processes,
which are compactly expressed with the traces of the products of the
Wilson line factors as follows:
\begin{eqnarray}
P_{\rm ss}&\equiv&
U_{\rm ss}(\bx,\by) U^*_{\rm ss}(\bar \bx,\bar \by) 
=
\frac{1}{N^2}
{\cal O}_1(\bx,\by,\bar \bx,\bar \by),
\end{eqnarray}
\begin{eqnarray}
P_{\rm as}&\equiv&
\sum_a U^a_{\rm as}(\bx,\by) U^{a*}_{\rm as}(\bar \bx,\bar \by) 
\nonumber \\
&=&
\frac{1}{N}{\cal O}_2(\bx,\by, \bar \bx, \bar \by)
-\frac{1}{N^2}{\cal O}_1(\bx,\by,\bar \bx,\bar \by),
\end{eqnarray}
\begin{eqnarray}
P_{\rm sa}&\equiv&
\sum_a U^a_{\rm sa}(\bx,\by) U^{a*}_{\rm sa}(\bar \bx,\bar \by) 
\nonumber \\
&=&
\frac{1}{N}{\cal O}_2(\bx,\bar \bx,\by, \bar \by)
-\frac{1}{N^2}{\cal O}_1(\bx,\by,\bar \bx,\bar \by),
\end{eqnarray}
\begin{eqnarray}
P_{\rm aa}&\equiv&
\sum_{a,b} U^{ba}_{\rm aa}(\bx,\by) U^{ba*}_{\rm aa}(\bar \bx,\bar \by) 
\nonumber \\
&=&
{\cal O}_1(\bx,\bar \bx,\by,\bar \by)
-\frac{1}{N}
{\cal O}_2(\bx,\by,\bar \bx,\bar \by)
\nonumber \\
&&
-\frac{1}{N}
{\cal O}_2(\bx,\bar \bx,\by,\bar \by)
+\frac{1}{N^2}{\cal O}_1(\bx,\by,\bar \bx,\bar \by),
\label{eq:sumsquare}
\end{eqnarray}
where we defined
\begin{eqnarray}
{\cal O}_1(\bx,\by,\bar \bx,\bar \by)&\equiv&
\Tr[W(\bx)W^\dagger(\by)]\ \Tr[W(\bar \by)W^\dagger(\bar \bx)]
,
\nonumber \\
{\cal O}_2(\bx,\by,\bar \bx,\bar \by)&\equiv&
\Tr[W(\bx)W^\dagger(\by)W(\bar \by)W^\dagger(\bar \bx)]
.
\label{eq:O}
\end{eqnarray}

\subsection{cross sections}

For the sake of simplicity, we assume here the nuclear target with
infinite transverse extent hypothetically.
The survival probability for a color singlet initial state,
$\varphi_0(\br,z)\xi_{\rm s}$,
is expressed as
\begin{eqnarray}
S&=&\frac{1}{\cal{A}}
      \int d\bx d\by dz d\bar \bx d\bar \by d\bar z
      \  \delta^2(\bR-\bar \bR)
\nonumber \\
&&\qquad\times
      \varphi_0^2(\br,z)\varphi_0^2(\bar \br,\bar z)
\overline{P_{\rm ss}(\bx,\by;\bar \bx, \bar \by)}
\nonumber \\
&=&  \int d\br d\bar \br
     \rho_0(\br)\rho_0(\bar \br)
\overline{P_{\rm ss}(\br,\bar \br, \bzero)}
\ ,
\label{eq:survival}
\end{eqnarray}
where $\cal A$ is the normalization transverse area and the $\delta$
function appears after the integration over $\bQ$. 
Here we also introduced 
$\rho_0(\br)\equiv \int dz \varphi_0^2(\br,z)$.  
Note that
$\overline {P_{\rm fi}}(\bx,\by;\bar \bx, \bar \by)=
\overline {P_{\rm fi}}(\br, \bar \br, \bR-\bar\bR)$
due to the assumption of the translation invariance 
in the transverse plane of the target.
Then the normalization area $\cal A$ is canceled out correctly.

At this point it may be useful to give the expressions for the
total and diffraction cross sections as well as the elastic cross
section of the singlet state.
The total cross section is obtained by taking the closure of all the
possible pair final states as
\begin{eqnarray}
\sigma^{\rm tot}_{\cal A}&=&
      \int d\bx d\by dz 
      \varphi_0^2(\br,z)
\nonumber \\
&&\quad \times
    \frac{1}{N}\overline{\Tr\left [(U(\bx,\by)-1)
          (U^\dagger(\bx,\by)-1)\right ]}
\nonumber \\
&=&
{\cal A} \int d\br \rho_0(\br)
   \left ( 2  - 2\Real  \overline{U_{\rm ss}}(\br) \right ),
\label{eq:total0}
\end{eqnarray}
where 
$\overline{U_{\rm ss}}(\bx,\by)=\overline{U_{\rm ss}}(\br)$,
and the suffix ${\cal A}$ indicates that this is the contribution
from the transverse area $\cal A$.
It is clear that
the factor 2=1+1 represents the incoherent sum of the contributions of the
incident and scattered waves, and the second term expresses the
interference.

The diffraction cross section, which is defined here as the processes
where the pair remains color-singlet without the net color exchange
with the target, but can be excited by the
interaction, is obtained by adding the color-singlet condition
on the out-going pair:
\begin{eqnarray}
\sigma^{\rm diff}_{\cal A}&=&
      \int d\bx d\by dz 
      \varphi_0^2(\br,z)
\nonumber \\
&&\quad \times
    \overline{(U_{\rm ss}(\bx,\by)-1)
          (U_{\rm ss}^*(\bx,\by)-1)}
\nonumber \\
&=&
{\cal A}   \int d\br \rho_0(\br)
   \left ( \overline{P_{\rm ss}}(\br, \br,\bzero)
  -  2\Real \overline{U_{\rm ss}}(\br) +1
\right )  
.
\end{eqnarray}
Finally the differential elastic cross section is
\begin{eqnarray}
&&\frac{d\sigma^{\rm el}_{\cal A}}{d\bQ/(2\pi)^2}
\nonumber \\
&=&
      \int d\bx d\by dz d\bar \bx d\bar \by d\bar z 
      \varphi_0^2(\br,z)     \varphi_0^2(\bar \br,\bar z)
      e^{i\bQ(\bR-\bar \bR)}
\nonumber \\
&&\quad \times    
    \overline{(U_{\rm ss}(\bx,\by)-1)
          (U_{\rm ss}^\dagger(\bar \bx,\bar \by)-1)}
\nonumber \\
&=&
{\cal A} \int d(\bR-\bar \bR) d\br d\bar \br 
      \rho_0(\br)\rho_0(\bar \br)  e^{i\bQ(\bR-\bar \bR)}
\nonumber \\
&&\times
   \left ( \overline{P_{\rm ss}}(\br,\bar \br,\bR-\bar \bR) 
  -   \overline{U_{\rm ss}}(\br)
  -   \overline{U^\dagger_{\rm ss}}(\bar \br)
 +1
\right ).
\end{eqnarray}
We note here that 
the inelastic cross section defined by
$\sigma^{\rm in}_{\cal A}=
\sigma^{\rm tot}_{\cal A}-\sigma^{\rm el}_{\cal A}$,
satisfies the usual relation
\begin{eqnarray}
\sigma^{\rm in}_{\cal A}
&=&
{\cal A} \int d\br d\bar \br 
      \rho_0(\br)\rho_0(\bar \br)  
   \left (1 - \overline{P_{\rm ss}}(\br,\bar \br,0)\right ) .
\nonumber \\
&=&
{\cal A} (1-S).
\end{eqnarray}

\section{Model of the target}
\label{sec:target}

We are considering the target as a source of the background gauge
fields created as the fluctuations in the nucleons in the thick 
target of length $L$.
We divide the path of the pair into $n$ different small zones.
The size $l$ (=$L/n$) of the zone may be taken as the correlation length
of the fluctuations, which should be of the order of, or smaller 
than the nucleon size in the target rest frame.
We assume here that the fields in the different zones are uncorrelated
and then we take the target average in each zone independently
\cite{BH96,H00,KW01}.
The eikonal factor 
is now factorized into a product of the contribution from each zone:
\begin{eqnarray}
W(\bx)=W(\bx)_{(n)} \equiv W_n(\bx) \cdots W_2(\bx) W_1(\bx).
\end{eqnarray}
We indicate here, and hereafter if necessary, 
by the lower suffix $(n)$ the fact that the quantity is a product of $n$ 
factors from the independent zones in the target.
In the transverse direction, we define 
the distribution of the color field fluctuations by 
\begin{equation}
\overline{A^a(\bx)A^b(\by)}\equiv \delta^{ab}C(\bx-\by)
\ ,
\label{eq:correlation}
\end{equation}
where $A^a= g \Delta x^+ A^{a-}$ with $\Delta x^+=\sqrt{2}l$.
Note that the transverse correlation length should be the same order
of $l$ for consistency.

We follow Ref.\cite{KW01} about the field averaging and
the details are presented in Appendix \ref{app1}.
We expect that the fluctuating field $A$ in each zone is weak and 
expand the eikonal amplitude up to $O(g^2)$.
For example\cite{H00,KW01},
the eikonal factor appearing in the total cross section
(\ref{eq:total0}), averages
\begin{eqnarray}
\overline{U_{\rm ss}}(\bx,\by)
&=&\frac{1}{N}\overline{\Tr(W(\bx) W^\dagger(\by))}_{(n)}
\nonumber \\
&=&
[1-C_F v(\br)]^n \simeq e^{-nC_F v(\br)}
\ ,
\end{eqnarray}
where $C_F=\frac{N^2-1}{2N}$ and
$v(\br)\equiv C(0)-C(\br)>0$
the subtracted correlation function. 

In the diffraction cross section, which yields the condition
$\bx=\bar \bx$ and $\by=\bar \by$ for the transverse positions in the
amplitudes and the conjugate one,
the eikonal factors 
$P_{\rm fi}(\bx, \by;\bx,\by)$ are reduced to
\begin{eqnarray}
P_{\rm ss}(\bx, \by;\bx,\by)&=&
|U_{\rm ss}(\bx,\by)|^2,
\nonumber \\
P_{\rm as}(\bx, \by;\bx,\by)&=&P_{\rm sa}(\bx, \by;\bx,\by)=
1-|U_{\rm ss}(\bx,\by)|^2,
\nonumber \\
P_{\rm aa}(\bx, \by;\bx,\by)&=&
N^2-2+|U_{\rm ss}(\bx,\by)|^2.
\label{eq:incl1}
\end{eqnarray}
The target average results in a compact form
\begin{eqnarray}
\overline{P_{\rm ss}}(\bx, \by;\bx,\by)
 &=& \frac{1}{N^2}\left [1+(N^2-1)(1-N v(\br))^n \right ]
\nonumber \\
&\simeq&\frac{1}{N^2}\left [
1+(N^2-1)e^{-n N v(\br)} \right ],
\label{eq:inclsol}
\end{eqnarray}
and those for other channels follows from the above relation
(\ref{eq:incl1}).
Eq.~(\ref{eq:inclsol}) is regarded as the probability that 
the singlet pair of size $\br$
remains singlet.
It shows that the equal-probability ($1/N^2$) distribution
in the color space is achieved exponentially in $L=nl$ for the quark pair 
with a definite size $\br$\cite{FM02}.
The result of  
\begin{eqnarray}
&&\overline{P_{\rm ss}}(\br, \br,\bzero)
  -  2\Real \overline{U_{\rm ss}}(\br) +1
\nonumber \\
&=&
\frac{1}{N^2}+1-2e^{-nC_Fv(\br)}+\frac{N^2-1}{N^2}e^{-nNv(\br)}
\end{eqnarray}
is already given in Ref.\cite{KW01}.

With this model of the target we obtain the recursion formula for the
average of the eikonal factor $\overline{P_{\rm fi}}(\bx,\by;\bar \bx,\bar \by)$
(see Appendix \ref{app1}), which is easily solved
as 
\begin{eqnarray}
\overline{ \left(
\begin{array}{cc}
P_{\rm  aa} & P_{\rm as} \\
P_{\rm sa} & P_{\rm ss} 
\end{array}
\right )}_{(n)} 
&=&
{\cal U} \ 
\overline{
\left (
\begin{array}{cc}
P_{\rm aa} & P_{\rm as} \\
P_{\rm sa} & P_{\rm ss} 
\end{array}
\right )}_{(n-1)} 
\nonumber \\
&=&{\cal U}^n\left (
\begin{array}{cc}
N^2-1 & 0 \\ 0 & 1 \end{array} \right )
   .
\label{eq:recursion}
\end{eqnarray}
The fact that the equation is closed in $\overline{ P_{\rm fi}}$ 
$({\rm f, i} = {\rm s, a})$ only,
 shows that
the transition between the singlet and adjoint 
states of the pair is a stochastic process during the travel 
through the target. That is, the non-vanishing contributions in the
color averaging come from those where the color states of the pair
in the amplitude and in the conjugate amplitude are the same in each step.

The explicit matrix elements of ${\cal U}$ are
\begin{eqnarray}
{\cal U}_{\rm aa} &=&
1-C_F[v(\bx-\bar \bx)+v(\by-\bar \by)]
\nonumber \\
&& +\frac{1}{2N}[v(\bx-\bar \bx)+ v(\by-\bar \by)
              +  v(\bx- \by)
\nonumber \\
&&\quad       +   v(\bar \bx-\bar \by)
              -2 v(\bx-\bar \by)-2v(\by-\bar \bx) ] ,
\nonumber \\
{\cal U}_{\rm as} 
&=& C_F [  v(\bx-\bar \by)+v(\by-\bar \bx) 
         - v(\bx-\bar \bx)-v(\by-\bar \by) ]
\ ,
\nonumber \\
{\cal U}_{\rm sa} &=&\frac{1}{N^2-1}\ {\cal U}_{\rm as}
\  ,
\nonumber \\
{\cal U}_{\rm ss} 
&=& 1-C_F [v(\bx- \by)+v(\bar \bx-\bar \by)]
.
\label{eq:elements}
\end{eqnarray}
As we summed over
the orientations of the adjoint vectors,
${\cal U}_{\rm as}$ is $N^2-1$ times as large as
${\cal U}_{\rm sa}$ 

Since ${\cal U}$ is a 2$\times$2 matrix, the calculation of 
${\cal U}^n$ is a simple task once the explicit form is given.
For a thick target (large $n$), ${\cal U}^n$ may be sharply
peaked at $\bx=\by=\bar \bx=\bar \by$, where ${\cal U}$ is unity,
and therefore we can approximate it
with a Gaussian in the sense of the steepest descent: The explicit
form of ${\cal U}$ we need in this approximation is
\begin{eqnarray}
{\cal U}_{\rm aa} &=&
1-\frac{\kappa}{4N}[4N^2 (\bR-\bar \bR)^2
\nonumber \\
&& 
\qquad +(N^2-3)(\br-\bar \br)^2+(\br+ \bar \br)^2],
\nonumber \\
{\cal U}_{\rm as} &=& 2C_F \kappa \br\cdot \bar \br,
\nonumber \\
{\cal U}_{\rm sa} &=&\frac{\kappa}{N}\br\cdot \bar \br ,
\nonumber \\
{\cal U}_{\rm ss} &=& 1-C_F\kappa(\br^2+\bar \br^2)
\label{eq:elements2}
\end{eqnarray}
with $v(\br) \simeq \kappa \br^2$ 
and $\kappa \propto \overline{(\nabla_\perp A^a(0))^2}$. 
We note again that the color transparency effects are properly included
in the fact that ${\cal U}_{\rm ss}=1$ and
${\cal U}_{\rm as}={\cal U}_{\rm sa}=0$ for $\br$ and/or $\bar \br=\bzero$.
From this expression for $\cal U$ we find explicitly that
$\overline{P_{\rm fi}}$ is a function of  
$\br, \bar \br, \bR-\bar \bR$ as noticed in the previous section
based on the translation
invariance of the target.
The $\bR-\bar \bR$ dependence appearing in ${\cal U}_{\rm aa}$ should
lead to the diffusion of the CM transverse momentum of the surviving
bound state.

\section{Singlet bound state penetration}
\label{sec:penetration}

In the thin target limit,
the interaction should be described with $n=1$, that is, single zone
on the path.
Noting that in the thin nuclear target
the inelastic cross section is a sum of those of the individual
nucleons in the target 
$\sigma^{\rm in}_{\cal A}={\cal A} l n_0 \sigma^{\rm in}$
with $n_0$ nuclear number density,
we can directly relate the expectation value 
of the correlation function $v(\br)$
to the inelastic cross section of the singlet bound state with a
nucleon through 
\begin{eqnarray}
({\cal A} l n_0)\ \sigma^{\rm in} 
&\equiv& \int d \bR d\br d \bar \br
\rho_0(\br) \rho_0(\bar \br)
(1 - {\cal U}_{\rm ss})
\nonumber \\
&=& {\cal A} \int d \br d \bar \br
\rho_0(r) \rho_0(\bar \br) C_F(v(\br)+v(\bar \br))
\nonumber \\
&=& 2 {\cal A} C_F \langle v \rangle
\end{eqnarray}
with $\langle v \rangle=\int d\br \rho_0(\br)v(\br)$.

Converting $\langle v \rangle$ with the physical quantity 
$\sigma^{\rm in}$,
we find that 
the survival probability (\ref{eq:survival}) for a small $L= n l $
has the usual form 
$S=1-\sigma^{\rm in} n_0 L= 1-L/L_{\rm in} 
\simeq e^{-L/L_{\rm in}}$
with $1/L_{\rm in}=\sigma^{\rm in}n_0
=2C_F \langle v \rangle l^{-1}$.

For a thick target the eikonal factor to calculate
the survival probability of the
singlet bound state is expressed as
\begin{eqnarray}
\overline{P_{\rm ss}}(\br, \bar \br, \bzero)
=
f_{0;1} e^{-n\ln(1/\lambda_1)} + f_{0;2} e^{-n\ln(1/\lambda_2)}
\label{eq:sskernel}
\end{eqnarray}
where $f_{0;1,2}$ are the singlet components squared of the eigenvectors
associated with the eigenvalues $\lambda_{1,2}$, respectively, 
of the matrix ${\cal U}$.
For a thick nuclear target to which the approximation 
(\ref{eq:elements2}) applies, the
explicit forms for these factors are found to be
\begin{eqnarray}
f_{0;1,2} &=& \frac{1}{2}\mp \frac{2 r_-^2+(N^2-2)r_+^2}
          {2N[4r_-^4+N^2r_+^4-4r_-^2 r_+^2]^{1/2}}
\label{eq:eigen1}
\ ,
\\
\lambda_{1,2}&=&1-\frac{\kappa}{8N} \Big [
2(N^2-2)r_-^2+N^2 r_+^2 
\nonumber \\
&&\qquad   \qquad  \left . 
\mp N
[4r_-^4+N^2r_+^4-4r_-^2 r_+^2 ]^{1/2} \right ]
\label{eq:eigen2}
\end{eqnarray}
with $r_\pm^2=(\br\pm\bar \br)^2$.
The eigenvalues satisfy $\lambda_2 \le \lambda_1 \le 1$, 
and  $\lambda_1=1$ when $\br=\bar \br$, which corresponds to the
unit eigenvalue of the probability matrix (\ref{eq:inclU})
in the diffraction.
With $N=2$ this reproduces the result of Ref.\cite{FM02}.

In the asymptotic region the larger eigenvalue 
$\lambda_1$ controls the behavior of the probability.
Noting that in the exponent
\begin{eqnarray}
\ln( 1/\lambda_1)\simeq \kappa C_F r_-^2  /2
\ ,
\end{eqnarray}
the asymptotic behavior of the survival probability is
obtained as
\begin{eqnarray}
S&\simeq&
     \int d\br d\bar \br
     \rho_0(\br)\rho_0(\bar \br)
     f_{0;1}e^{-n\kappa C_F r_-^2 / 2}
\nonumber \\
&\simeq&
     \frac{1}{N^2} 
     \int d\br 
     \rho_0(\br)^2 
     \int d \br_- e^{-n\kappa C_F r_-^2 / 2}
\nonumber \\
&\simeq&
\frac{1}{N^2}\frac{\Delta}{n\kappa C_F}
\ ,
\label{eq:asymp1}
\end{eqnarray}
where $\Delta = 2\pi \int d\br \rho_0^2(\br)$. In SU(2) the
fundamental representation is pseudo-real and 
Eqs.~(\ref{eq:eigen1}) and (\ref{eq:eigen2}) become 
symmetric in $r_+^2 \leftrightarrow r_-^2$ with $N=2$,
which multiplies an extra factor of 2 on RHS\cite{FM02}.
The most important fact we observe here
is that the integration over $\br$ with
the weight $\rho_0(\br)$ changes the superficial exponential dependence 
on $n\propto L$ into the power dependence, $1/n \propto 1/L$.

For a small bound state
where $\langle v \rangle \simeq \kappa \langle \br^2 \rangle$
is valid even for a thin target, the 
single parameter $\kappa$ determines the behavior of the
survival probability from the thin to the thick
target region. With
$L_{\rm in}$ the asymptotic form (\ref{eq:asymp1}) is rewritten as
\begin{equation}
S \simeq \frac{\Delta}{N^2}\ 
   \frac{1}{2\pi}
   \int d\br_- e^{-L r_-^2 /4L_{\rm in}\langle \br^2 \rangle}
=2 \ \frac{\Delta \langle \br^2 \rangle}{N^2}\frac{ L_{\rm in}}{L} . 
\label{eq:asymp2}
\end{equation} 
In this form the $N$ dependence appears in the denominator only, which
comes from the equi-partitioning in the color space.

It would be suggestive to show an explicit calculation with the
Gaussian wavefunction 
$\rho_{0}(r)=\frac{1}{\pi\langle \br^2\rangle}
e^{-r^2/\langle \br^2 \rangle}$
and using the form $v(\br)=\kappa \br^2$. For the Gaussian wave
function $\Delta \langle \br^2 \rangle =1$.
The survival probability scales as $S=S(L/L_{\rm in})$ 
since the exponent in the eikonal kernel is quadratic
in $\br$ and $\bar \br$ and we can rescale the integration variable to
$\br/\langle \br^2 \rangle^{1/2}$ 
and $\bar \br/\langle \br^2 \rangle ^{1/2}$
(see Eqs.~(\ref{eq:eigen1}), (\ref{eq:eigen2})),
which results in a prefactor, $n \kappa \langle \br^2\rangle$, 
in the exponent of the eikonal kernel.

\begin{figure}[tb]
\begin{center}
\epsfxsize=0.4\textwidth
\epsffile{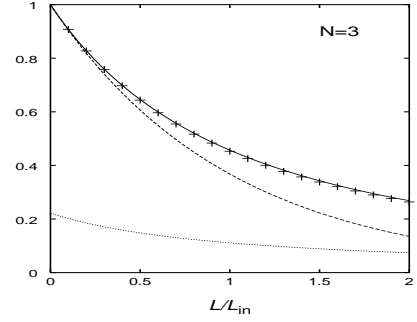}
\caption{Survival probability of the color singlet Gaussian bound
  state passing through SU(3) random background fields. The
analytic form (\protect\ref{eq:analform}) [solid] and the simple exponential
form [dashed] are compared with the numerical result [cross].
The first term of Eq.~(\protect\ref{eq:analform}) is also shown in dotted line.
\label{fig:2}
}
\end{center}
\end{figure}

In U(1) case this simply yields
\begin{eqnarray}
S
&=& 
  \frac{1}{1+2n\kappa\langle \br^2 \rangle}
=
  \frac{1}{1+L/L_{\rm in}}
\  ,
\label{eq:penet_ho}
\end{eqnarray}
where $L_{\rm in}\equiv 1/(2\kappa l^{-1}\langle \br^2 \rangle)$.
Eq.~(\ref{eq:penet_ho}) smoothly interpolates between the usual
absorption rate $1-L/L_{\rm in}$ for small $L$ and
$L_{\rm in}/L$ for large  $L$.

In SU($N$) we know the asymptotic form both in the large and small
$L$ regions from the above results in this section.
Furthermore, 
the integral (\ref{eq:survival}) with (\ref{eq:sskernel})
 in the region except for
$\br \simeq \bar \br$ should behave as $1/L^2$ for large $L$
since the exponent of the eikonal kernel is proportional to the
target width $L$.
From this consideration and motivated by the U(1) result we speculate
an analytic form for $S$ as follows:
\begin{eqnarray}
S(L/L_{\rm in})= \frac{2}{N^2} \frac{1}{1+\frac{L}{L_{\rm in}}}
+\frac{N^2-2}{N^2} \frac{1}{(1+\frac{1}{2}\frac{L}{L_{\rm in}})^2}.
\label{eq:analform}
\end{eqnarray}
This reproduces the correct small  and large $L$ behavior of $S$.

We compare this form (\ref{eq:analform}) 
with the result of the numerical integration in
Fig.~\ref{fig:2}. We find that the simple analytic form
(\ref{eq:analform}) approximates the numerical result very well.
We confirmed this for $N=3\sim10$ and $L/L_{\rm in}=0\sim10$.
The fit in SU(2) is not as precise
as in SU($N\ge 3$), due to the extra factor of 2 in the
asymptotic region for SU(2), but remains very much reasonable 
in the region of $L/L_{\rm in}=0\sim2$.

In summary the survival probability of the singlet bound state
decays in 1/$L$ in the asymptotically large $L$. However, the approach
to this asymptotic behavior is rather slow and
the second term in Eq.~(\ref{eq:analform}) equally contributes
in the intermediate region of $L/L_{\rm in}$. For the larger $N$,
it can even dominate over the first term. 
Nevertheless, the suppression of the survival probability is given
in the power of $1/L$, as typically shown in
Eq.~(\ref{eq:analform}), 
not in the exponential one.

\section{Discussion}
\label{sec:discussion}

Within the same approximation of 
$\langle v \rangle = \kappa \langle \br^2 \rangle$ and 
with the use of the Gaussian wavefunction,
the asymptotic forms for the total and diffraction cross sections are easily
found as 
\begin{eqnarray}
\sigma_{\cal A}^{\rm tot}&\simeq& {\cal A}
\int d\br \rho_0(\br)\left (
2-2e^{-L\br^2/2L_{\rm in}\langle \br^2 \rangle} 
\right ) 
\nonumber \\
&=&
2{\cal A}\left (1-\frac{1}{1+\frac{1}{2}\frac{L}{L_{\rm in}}}\right ) 
\ ,
\label{eq:total}
\end{eqnarray}
and 
\begin{eqnarray}
\sigma_{\cal A}^{\rm diff}
&\simeq& {\cal A}
\int d\br \rho_0(\br)\left (
\frac{1}{N^2}+1
 \right .
\nonumber \\
&&
\left .
+\frac{2C_F}{N}
e^{-\frac{N}{2C_F}\frac{L}{L_{\rm in}}\frac{\br^2}{\langle \br^2 \rangle}}
-2 e^{-\frac{L}{2L_{\rm in}}\frac{\br^2}{\langle \br^2 \rangle}}
\right ) 
\nonumber \\
&=&
{\cal A}\left [\frac{1}{N^2}+1
+\frac{2C_F}{N}\frac{1}{1+\frac{N}{2C_F}\frac{L}{L_{\rm in}}}-
\frac{2}{1+\frac{1}{2} \frac{L}{L_{\rm in}}} \right ].
\nonumber \\
\label{eq:diffraction}
\end{eqnarray}
We note that the above results show that
$\sigma^{\rm tot}_{\cal A}={\cal A}\frac{L}{L_{\rm in}}$ and
$\sigma^{\rm diff}_{\cal A}=O((\frac{L}{L_{\rm in}})^2)$ for small $L$. 
The factor $\frac{\br^2}{L_{\rm in}\langle \br^2 \rangle}$
in the exponent of the integrand, 
or more generally
\begin{equation}
\frac{v(\br)}{L_{\rm in}\langle v \rangle}
=
\frac{v(\br^2)}{\langle v \rangle}\sigma^{\rm in}n_0
\equiv
\sigma^{\rm in}(\br^2)n_0,
\end{equation}
might be interpreted
as a size-dependent cross section of the rigid dipole of size~$\br$.
This fact that the smaller component is less interactive
in passing through the target is usually referred
to as {\it color transparency}.
The integration over the incoherent contributions of the color dipoles
with size $\br$ leads to the power behavior in $L$ for the total
and diffractive cross sections\cite{Ba96}. 
In the total cross section (\ref{eq:total})
the interference term between the initial
and scattered waves decays in $1/L$, approaching the black-disk
limit $\sigma^{\rm tot}_{\rm BD}=2{\cal A}$. 
The diffraction cross section also approaches 
the incoherent sum of the incident and scattered wave contributions,
${\cal A}(1+\frac{1}{N^2})$, the latter of 
which is suppressed by the equi-partitioning factor $\frac{1}{N^2}$
in the color space. 

The asymptotic $L^{-1}$-dependence of the survival probability
looks apparently the
same as the one seen in the total cross section.
The exponent of the integrand in Eq.~(\ref{eq:sskernel}), however, 
no longer has the simple meaning of the cross section 
$\sigma^{\rm in}(\br^2)$ of the dipole 
of size $\br$,
but depends on both of
$\br$ and $\bar \br$ in the amplitudes. 
Actually there are two exponential factors in the eikonal kernel;
$\lambda_1$ which controls the asymptotic behavior of $S$, weights the
diagonal components ($\br \simeq \bar \br$) of the amplitudes, 
and the other expresses the color diffusion in the color space.

The question about the energy scale
where the change of the survival probability
from the usual exponential form to the power form
occurs should be addressed. For this purpose one would have to includes
the longitudinal momentum transfers and the binding energies of the
resonances. We leave this extension of the framework for future study.

Another energy dependence will come from the splitting of the
quark to a quark and a gluon or evolution of the 
the quark pair wavefunction\cite{M94,KW01}. 
Such an evolution may be also necessary to our model 
at very high energies.

In the infinite momentum frame of the target, on the other hand,
this branching should be included in the target wavefunction.
In McLerran-Venugopalan model~\cite{MV94} the valence 
charge distribution is treated as 
the random variable in the longitudinal as well as the  transverse
directions of the target.
The generated classical fields
has the subtracted correlation
$v(\br)=\kappa \br^2 \ln (1/\br^2 \Lambda^2)$
for sufficiently small $\br^2\ll \Lambda^2$
with $\Lambda$  the infrared cutoff, 
reflecting the massless nature of the fields.
The inclusion of the 
quantum corrections to this model has been intensively
studied at very high energies (see e.g., \cite{RG}).
This energy dependence might be economically mimicked by using
the energy-dependent correlation function $v(\br)$ in our
framework.

As for the J/$\psi$ production in heavy ion collisions (HIC)
we need the initial quark
pair wavefunction whose size is typically of the order $1/2m_c$.
At very high energy there is no time for the quark pair to expand,
and the quarks stay
close to each other during two nuclei going through. In this
situation, the color fields in the nuclei will affect
the penetration of the singlet state only weakly due to 
the color transparency. 
From the same reason,
the transitions between the singlet and the octet states will be
suppressed as seen in Eq.~(\ref{eq:elements2})\cite{Matsui}.

One can argue based on the eikonal factor
(\ref{eq:elements2})
that the CM momentum of the octet pair will diffuse stronger
than that of the singlet one. 
The difference in the CM momentum dependence
between the singlet and adjoint states may give a hint on the
production mechanism of the J/$\psi$ in the HIC experiments.

Finally we notice that the J/$\psi$ 
production rate should given by 
the square of the total amplitude. Here we concentrated only
on the interaction of the $Q\bar Q$ pair with
the classical color fields in the target nucleus, ignoring other effects;
for instance, the action of the background gauge fields on the
quark pair may induce the gluon emissions, which will be important for
the energy-loss and the color neutralization of the pair. We also leave 
these effects for future study.

\section{summary}
\label{sec:summary}

In this paper we studied the absorption of the
high-energy quark--anti-quark bound state in the random, classical,
background SU($N$) gauge fields. 
Using the eikonal approximation for the
quark and anti-quark propagations, we found that the probability is
suppressed by the inverse power of the thickness $L$ of the target
for general SU($N$). 
The asymptotic form,
$\frac{2}{N^2}\frac{L_{\rm in}}{L}$,
was given, but its approach is rather slow in $L$.
For the small Gaussian bound state, 
we interpolated between the known asymptotic forms 
in the small and large $L$ regions by 
the rational form (\ref{eq:analform}), and found that it
gives a remarkably precise
description for the numerical result 
in the whole $L/L_{\rm in}$ region.
At very high energies the survival probability
cannot have the exponential suppression behavior.

In the inclusive cases such as the total and the diffractive cross
sections, the cross section of the bound state may be regarded as
a sum of the dipole components of definite size $\br$.
 However in more exclusive cases, e.g.,
the penetration probability $S$, this simple interpretation does not
apply.

The survival probability of the J/$\psi$ 
has an importance in the HIC physics as a QGP sensitive probe.
In order to investigate the issue of the J/$\psi$ yield in HIC
more realistically, 
we must certainly extend our model to include the processes of
the heavy-pair production,
the interactions with other hard constituents in the hot matter, and
so on.

\acknowledgements

The author is very thankful to T.~Matsui for fruitful discussions
on this subject, and is grateful to 
D.~Kharzeev, L.~McLerran and R.~Venugopalan
for helpful conversations.
He acknowledges the hospitalities of RBRC at BNL and YITP at Kyoto U.,
where parts of this work were done.
This work is supported in part by the Grants-in-Aid for Scientific Research of
Monbu-kagaku-sho, No.\ 1340067.

\appendix
\section{Target Averages}
\label{app1}

We follow Ref.\cite{KW01} in the target average.
To perform the target average for the last zone
we expand the eikonal factor or the Wilson line up to $O(g^2)$
\begin{eqnarray}
&&W(\bx)_{(n)}= 
W_n(\bx)W(\bx)_{(n-1)}
\nonumber \\
&=&
\left (1+i A^a(\bx,n) t^a - \frac{C_F}{2} C(\bzero) \right )
W(\bx)_{(n-1)}.
\end{eqnarray} 
The target average is defined by
\begin{equation}
\overline{A^a(\bx,n)A^b(\by,n)}=\delta^{ab}C(\bx-\by)
\end{equation}
and we utilize an elementary relation 
$t^a_{ij} t^a_{kl}=
\frac{1}{2}(\delta_{il}\delta_{jk} -
\frac{1}{N}\delta_{ij}\delta_{kl})$ for the fundamental representation.
In the above expression we have averaged the $O(g^2)$ term as
we are working only up to this order.
With this definition for the target average,
we obtain
\begin{eqnarray}
\overline{U_{\rm ss}}(\bx,\by)
&=&\frac{1}{N}\overline{\Tr(W(\bx) W^\dagger(\by))}_{(n)}
\nonumber \\
&=&
\frac{1}{N}\overline{\Tr(W(\bx)W^\dagger(\by) )}_{(n-1)}
[1-C_F v(\br)] 
\nonumber \\
&=&
[1-C_F v(\br)]^n \simeq e^{-nC_F v(\br)}.
\end{eqnarray}

The target average for the eikonal kernel for the singlet
survival probability $P_{\rm ss}$ is expressed as a trace of 
four Wilson line factors. First let us perform the target average of
${\cal O}_1$ and ${\cal O}_2$ given in Eq.~(\ref{eq:O}).
By taking the average for the last zone we find
\begin{eqnarray}
&&
\overline{\left (
\begin{array}{c} {\cal O}_1 \\ {\cal O}_2 \end{array} 
\right )}_{(n)}
(\bx,\by,\bar \bx,\bar \by)
\nonumber \\
&=&
\left ( \begin{array}{cc}
\alpha-\frac{1}{2N}\beta & \frac{1}{2}\beta \\
\frac{1}{2}\beta'        & \alpha'-\frac{1}{2N}\beta'
\end{array} \right )
\overline{\left (
\begin{array}{c} {\cal O}_1 \\ {\cal O}_2 \end{array} 
\right )}_{(n-1)}(\bx,\by,\bar \bx,\bar \by)
 ,
\end{eqnarray}
where 
\begin{eqnarray}
\alpha &=&1-C_F(v(\bx-\by)+v(\bar \bx-\bar \by) )
\nonumber \\
\beta  &=& v(\bx-\bar \by)+v(\bar \bx- \by)-v(\bx-\bar \bx)-v(\by-\bar \by)
\nonumber \\
\alpha' &=&1-C_F(v(\bx-\bar \bx)+v(\by-\bar \by) )
\nonumber \\
\beta'  &=& v(\bx-\bar \by)+v(\bar \bx- \by)-v(\bx-\by)-v(\bar \bx-\bar \by)
.
\end{eqnarray}

Making the physical combinations, 
$P_{\rm fi}(\bx,\by;\bar \bx,\bar \by)$, given in
Eq.(\ref{eq:sumsquare}), we find a closed relation 
\begin{eqnarray}
&&\overline{\left (
\begin{array}{cc} 
P_{\rm aa }  & P_{\rm as }
\\
P_{\rm sa }  & P_{\rm ss }
\end{array} 
\right )}_{(n)}
\nonumber \\
&=&
\left ( \begin{array}{cc}
\alpha'-\frac{1}{2N}(\beta+\beta') &C_F \beta \\
\frac{1}{2N}\beta        & \alpha
\end{array} \right )
\overline{\left (
\begin{array}{cc} 
P_{\rm aa }  & P_{\rm as }
\\
P_{\rm sa }  & P_{\rm ss }
\end{array} 
\right )}_{(n-1)}
 ,
\end{eqnarray}
which is given in Eqs.~(\ref{eq:recursion}), (\ref{eq:elements}). 
The consistency may be
checked by taking the average with $n=1$ on LHS and noting that 
${ P_{\rm aa}}_{(0)}=N^2-1$,
${ P_{\rm as}}_{(0)}={P_{\rm sa}}_{(0)}=0$,
and ${P_{\rm ss}}_{(0)}=1$ on RHS.

The target average of the eikonal factors $P_{\rm fi}(\bx,\by;\bx,\by)$,
which appears in the diffraction case with $\rm f=i=s$,
are easier to do.
Actually $P_{\rm ss}(\bx,\by;\bx,\by)$ is a single independent element
as shown in Eq.~(\ref{eq:incl1}), and
its target average satisfies the relation
\begin{equation}
\overline{P_{\rm ss}}(\br,\br,\bzero)_{(n)}=
(1-Nv(\br))\overline{P_{\rm ss}}(\br,\br,\bzero)_{(n-1)}+\frac{1}{N}v(\br) ,
\end{equation}
which immediately gives the result Eq.~(\ref{eq:inclsol}).
The same is obtained by setting $\bx=\bar \bx$ and
$\by=\bar \by$ in the matrix, 
\begin{eqnarray}
{\cal U}=\left (\begin{array}{cc}
1-\frac{v(\br)}{N} &  (N^2-1)\frac{v(\br)}{N} \\
  \frac{v(\br)}{N} &1-(N^2-1)\frac{v(\br)}{N}
\end{array}
\right )
,
\label{eq:inclU}
\end{eqnarray}
which has now the property
of the probability matrix. The result yields
\begin{eqnarray}
&&\overline{P(\br,\br,\bzero)}_{(n)}
={\cal U}^n P_{(0)}
\nonumber \\
&=&\frac{1}{N^2}\left (\begin{array}{c}
(N^2-1)(N^2-1+(1-Nv(\br))^n)  \\ 
(N^2-1)(1-(1-Nv(\br))^n)    
\end{array} \right .
\nonumber \\
&& \left . \qquad \qquad \qquad
\begin{array}{c}
 (N^2-1)[1-(1-Nv(\br))^n ] \\
 1+(N^2-1)(1-Nv(\br))^n      
\end{array}
\right )
.
\end{eqnarray}

\section{More on SU(2)}
\label{app2}

As the fundamental representation of SU(2) is pseudo-real: 
$\tau^2 \tau^a \tau^2 =-\tau^{a*}$ with
Pauli matrices $\tau^a$, one finds for the Wilson lines
\begin{eqnarray}
\tau^2 W(\bx) \tau^2&=&W^\dagger (\bx),
\\
\Tr(W(\bx)W^\dagger(\by))
&=&\Tr(\tau_2 W(\bx)\tau_2 \tau_2 W^\dagger(\by) \tau_2)
\nonumber \\
&=& \frac{1}{N} \Tr(W(\bx)^\dagger  W(\by)),
\end{eqnarray}
and therefore for the eikonal factor of the singlet pair
\begin{eqnarray}
\overline{U_{\rm ss}}(\br)&=&
\overline{U_{\rm ss}}(-\br)
,
\\
\overline{U_{\rm ss}U_{\rm ss}^*}(\br,\bar \br,\bzero)&=&
\overline{U_{\rm ss}U_{\rm ss}^*}(\pm\br,\mp\bar \br,\bzero)
.
\end{eqnarray}

In Ref.~\cite{HLN90} a model of SU(2)
with the finite, fixed field magnitude $A$ is studied, 
where the importance of the color degrees of freedom in the color
dipole interactions at high energies 
is stressed and the spatial degrees are
omitted.
That model is hence similar to the diffraction case in our
framework with identifying the eikonal factor for each zone as
$W_{qi}=\cos A +2i t^a \hat A^a \sin A$ for the quark
and  $W_{\bar qi}^\dagger =\cos A -2i t^a \hat A'^a \sin A$
for the anti-quark.
The orientation $\hat A$ of the field $A^a$ is assumed random and the
average is accordingly defined by
$\overline{A^a  A^b}=\frac{1}{3}A^2 \delta ^{ab}$.
If assuming that the orientations of $\hat A$ and $\hat A'$ are
uncorrelated and random, one finds a recursion relation 
\begin{eqnarray}
\overline{P_{\rm ss}^{\rm uncor}}_{(n)}
&=&
(1-\frac{4}{3} \sin^2 A)^2
\overline{P_{\rm ss}^{\rm uncor}}_{(n-1)}
\nonumber \\
&& +\frac{2}{3}\sin^2 A (1-\frac{2}{3}\sin^2 A).
\end{eqnarray}
The solution is 
\begin{eqnarray}
\overline{P_{\rm ss}^{\rm uncor}}
=\frac{1}{4}\left [1+3 \left (
\frac{1+2 \cos 2A}{3} \right )^{2n} \right ], 
\label{eq:hufner1}
\end{eqnarray}
which is the result of Ref.\cite{HLN90}.
On the other hand, the case of $\hat A = - \hat A'$ where
the quark and the anti-quark see the correlated field,
which is actually simpler, one obtains
\begin{eqnarray}
\overline{P_{\rm ss}^{\rm corr}}_{(n)}&=&
(1-\frac{4}{3} \sin^2 2A)\overline{P_{\rm ss}^{\rm corr}}_{(n-1)}
+\frac{1}{3}\sin^2 2A
\\
&=&\frac{1}{4}\left [1+3 \left (
\frac{1+2 \cos 4A}{3} \right )^n \right ]. 
\label{eq:hufner2}
\end{eqnarray}
The $n$- and $N$-dependences of the results in both cases
are completely consistent with Eq.~(\ref{eq:inclsol}) with $N=2$.

\end{document}